\documentclass[%
 preprint,
groupedaddress,
 amsmath,amssymb,
aps,
]{revtex4-1}

\usepackage{hyperref}
\usepackage{graphicx}
\graphicspath{ {./} }
\usepackage{dcolumn}
\usepackage{bm}
\usepackage{epsfig}
\usepackage{epstopdf}
\usepackage{textcomp}
\usepackage{caption}
\usepackage{subcaption}
\usepackage{fullpage}
\usepackage{color}
\usepackage{epsfig}
\usepackage{epstopdf}
\usepackage{textcomp}
\usepackage{multirow}
\usepackage{setspace}
\usepackage{fullpage}
\usepackage{float}

\begin{document}

\title{Influence of interphase anisotropy on lamellar eutectic growth patterns}

\author{Supriyo Ghosh}
\affiliation{Condensed Matter Physics, Ecole Polytechnique, CNRS, 91128 Palaiseau, France}
\author{Mathis Plapp}
\affiliation{Condensed Matter Physics, Ecole Polytechnique, CNRS, 91128 Palaiseau, France}

\date{\today}

\begin{abstract}
It is well documented in many experiments that crystallographic 
effects play an important role in the generation of two-phase 
patterns during the solidification of eutectic alloys. In 
particular, in lamellar composites, large patches of perfectly 
aligned lamellae are frequently observed. Moreover, the growth 
direction of the lamellae often markedly differs from the 
direction of the temperature gradient (the lamellae are tilted 
with respect to the main growth direction). Both of these effects 
cannot be explained either by the standard theory or the available 
numerical models of eutectic growth, which all assume the interfaces 
to be isotropic. We have developed a phase-field model in which 
the anisotropy of each interface (solid-liquid and solid-solid) 
can be separately controlled, and we have investigated the effect 
of interface anisotropy on the growth dynamics. We have found that 
anisotropy of the solid-solid interphase boundary free energy 
dramatically alters the growth dynamics. Tilted lamellae result 
from the modified equilibrium condition at the triple lines, in 
good agreement with a theoretical conjecture proposed recently. 
In three dimensions, the interphase boundaries tend to align 
with directions of minimal energy. We have also performed 
simulations in which two grains with different anisotropies 
are in competition. In all cases, the grain containing the 
boundaries with the lowest energies was selected after a 
transient. These results shed new light on the selection of 
growth patterns in eutectic solidification.
\end{abstract}

\pacs{Valid PACS appear here}

\maketitle

\section{Introduction}
During directional solidification, a sample is pulled with a constant 
velocity $V$ in a fixed thermal gradient $G$ aligned with the $z$ 
direction. For a non-facted binary eutectic alloy, this leads to the
formation of composite materials: the solid consists of two phases
$\alpha$ and $\beta$, which grow together by diffusive exchange of 
components through the liquid~\cite{Jackson66}. In the absence of interfacial anisotropy, 
the solid phases grow next to each other, the interphase boundaries 
(interfaces between $\alpha$ and $\beta$) are aligned with the growth 
direction, and Young's law is satisfied at the trijunction points. 
Here, we investigate several phenomena that arise when the interphase
boundaries (IB) are anisotropic.

Whereas, in non-faceted substances, the solid-liquid surface free
energy is only weakly anisotropic, the anisotropy of solid-solid 
IB can be strong. This anisotropy depends on the relative
orientation of the two solids with respect to each other. The
anisotropy function thus is not an intrinsic property of the
material, but differs between different {\em eutectic grains}~\cite{Hogan71} --
that is, portions of the solid in which the orientations of the 
two phases remain homogeneous. One may distinguish {\em floating
grains} with low anisotropy from {\em locked grains} with high
anisotropy~\cite{Caroli92}. In the latter, the IB may remain ``locked'' onto a 
direction of low energy, irrespective of the orientation of the
grain with respect to the temperature gradient.

For our studies, we have used the grand-canonical phase-field model
described in Refs.~\cite{abhik2012,Ghosh15}. In this model, which
provides quantitative results due to a well-controlled thin-interface
limit \cite{abhik2012}, the anisotropy of each interface can be
chosen independently \cite{Ghosh15}. We write the surface energy 
$\gamma_{\alpha\beta}$ as
\begin{equation}
\gamma_{\alpha\beta}= \bar\gamma_{\alpha\beta} a_c(\theta,\phi),
\end{equation}
where $\bar\gamma_{\alpha\beta}$ is a constant which we take equal
to the solid-liquid interface free energies, and $a_c(\theta,\phi)$
is the dimensionless anisotropy function. Here, $\theta$ and $\phi$
are the standard polar angles with respect to the $z$ and $x$ axis,
respectively, and we take the convention that $\theta=\phi=0$ corresponds
to a minimum of $a_c$ (that is, a minimum in the interface energy
occurs when the boundary is in the $xz$ plane). In the so-called
rotating directional solidification set up \cite{silvere2012e}, the
orientation of the eutectic crystal with respect to the temperature
gradient can be changed during the experiment. We describe such
rotations around the $y$ and $z$ axes by the angles $\theta_R$ and
$\phi_R$, respectively.

We consider a generic binary eutectic alloy of symmetric phase diagram,
with equal volume fractions of the two solids. In this situation, the
capillary and thermal lengths associated with the two solids are equal.

\begin{figure}[h]
\begin{center}
\begin{subfigure}[b]{0.5\textwidth}
\centering
\includegraphics[width=\textwidth]{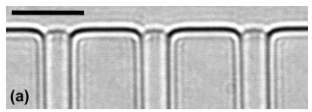}
\end{subfigure}%
\begin{subfigure}[b]{0.5\textwidth}
\centering
\includegraphics[width=\textwidth]{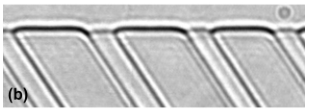}
\end{subfigure}
\caption{\label{fig1}Snapshots of directional solidification in thin
samples of the transparent alloy CBr$_4$--C$_2$Cl$_6$ (adapted from Akamatsu et al.~\cite{silvere2012t}); Bar: 20 $\mu$m.}
\end{center}
\end{figure}

\section{Tilted lamellae}
As shown in Fig.~\ref{fig1}b, lamellae do not always grow aligned
with the temperature gradient (which is vertical in the figure).
The growth angle results from a competition between the external 
temperature field and the interface anisotropy. Recently, a
conjecture was made that provides a prediction for the growth 
angle \cite{silvere2012t}. It is based on the observation 
(see Fig.~\ref{fig1}) that the solid-liquid interfaces 
(the ``lamellae heads'') remain symmetric even when the lamellae 
are tilted. This implies that the Cahn-Hoffman surface tension
vector ~\cite{Hoffman72} must be aligned with the $z$ axis (see
Refs.~\cite{silvere2012t,Ghosh15} for details). From this hypothesis,
one obtains
\begin{equation}
\gamma_{\alpha\beta}(\theta-\theta_R) \sin\theta + \gamma'_{\alpha\beta}(\theta-\theta_R)\cos\theta = 0,
\label{eq:SP}
\end{equation}
where the prime denotes differentiation with respect to $\theta$.
For a fixed orientation $\theta_R$ of the eutectic grain, this is a 
nonlinear equation for the interface orientation $\theta$, which 
can easily be solved numerically for arbitrary anisotropy functions 
$a_c(\theta)$. As long as the interface stiffness 
$\gamma+\gamma''$ is positive for all angles, this equation
has a unique solution. For negative stiffness, there are ranges of
$\theta_R$ for which there exist three solutions, of which one
corresponds to an orientation that is present on the equilibrium
shape (stable solution), one to an unstable orientation, and the third 
to a metastable orientation. This is the prediction against which we
will compare our numerical results.

\begin{figure}[h]
\begin{center}
\begin{subfigure}[b]{0.5\textwidth}
\centering
\includegraphics[width=\textwidth]{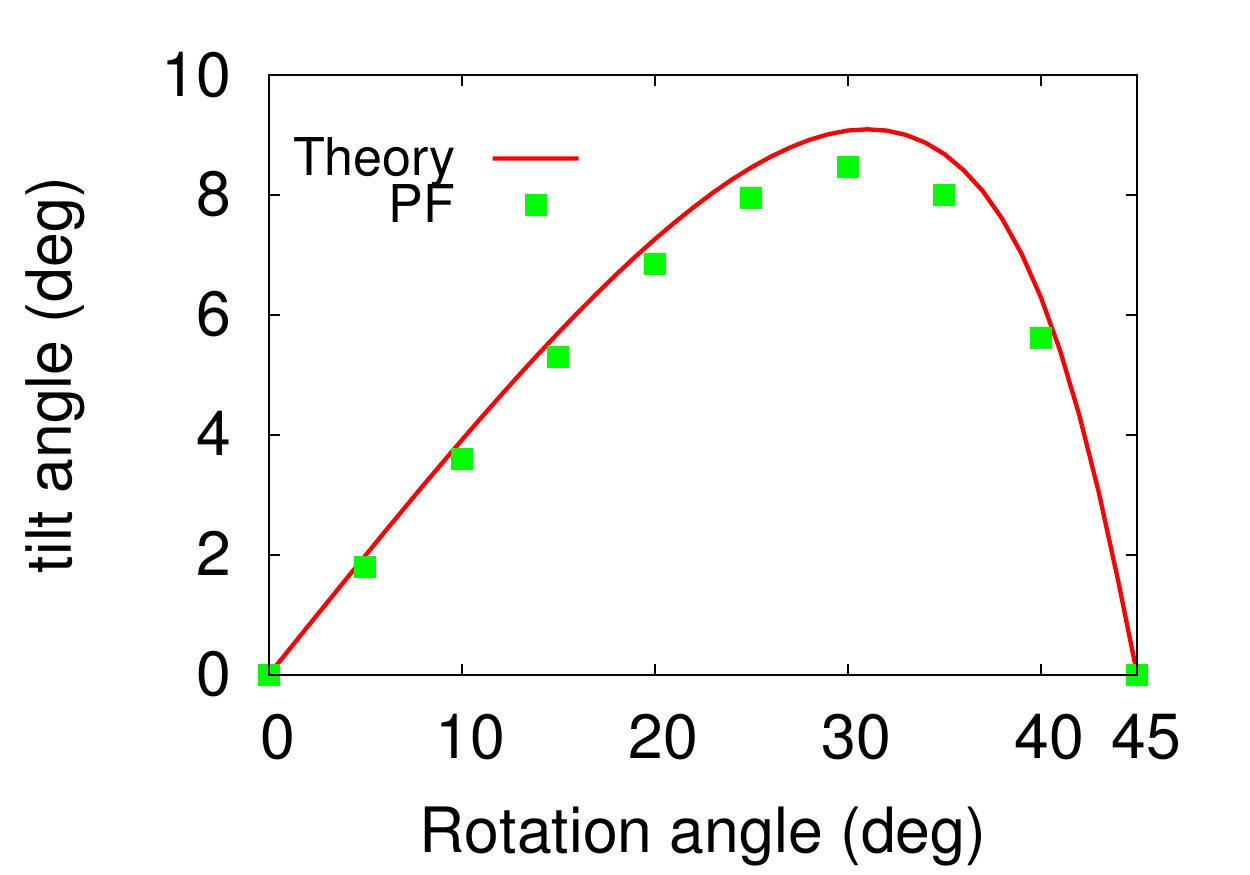}
\end{subfigure}%
\begin{subfigure}[b]{0.5\textwidth}
\centering
\includegraphics[width=\textwidth]{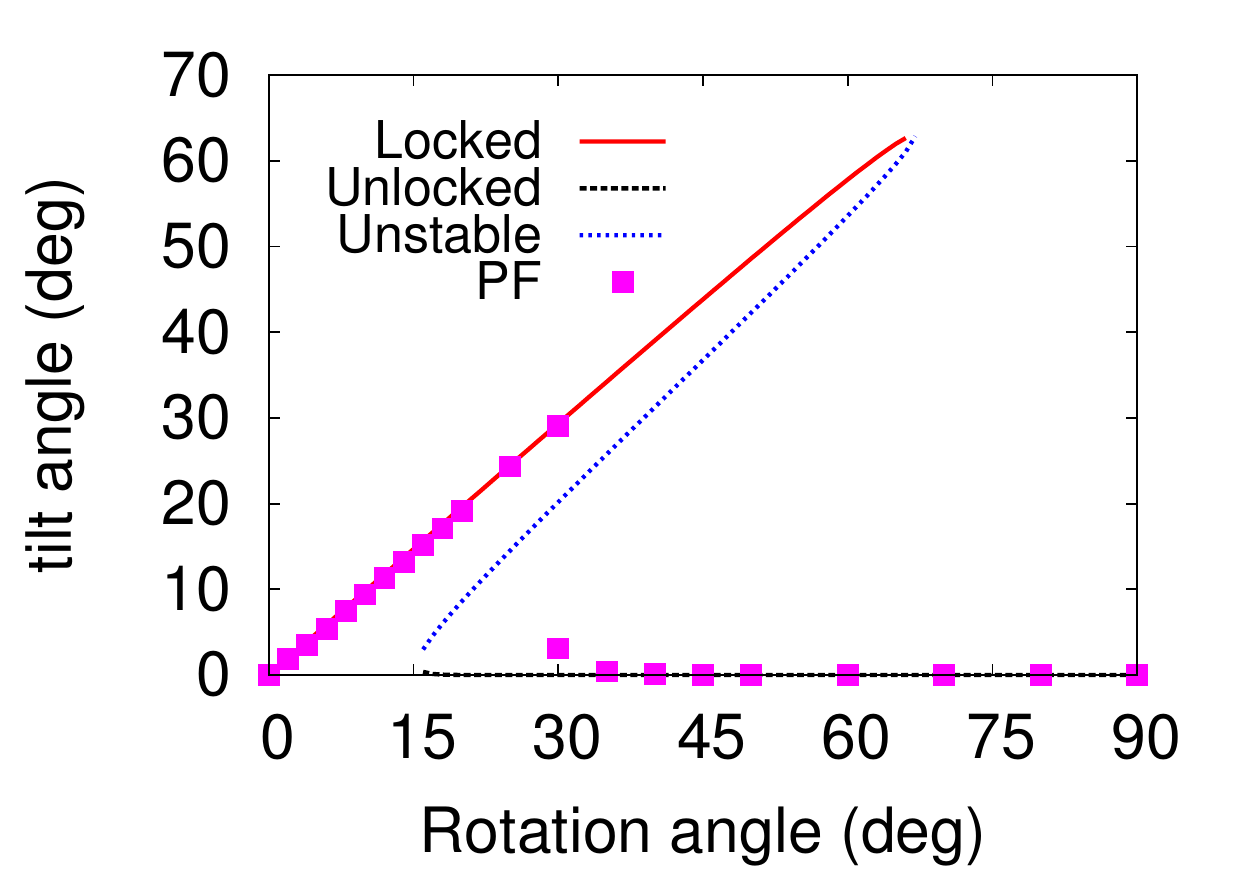}
\end{subfigure}
\caption{Interphase tilt angle for (left) smooth anisotropy function ($\epsilon_4 = 4\%$) (right) cusp anisotropy function with $\epsilon_c = 20\%$. The solid lines indicate the tilt angle predicted by Eq.~\protect\ref{eq:SP}; squares denotes the phase-field (PF) results.\label{fig_cusp}}
\end{center}
\end{figure}

We have performed simulations in two dimensions for various
anisotropy functions, with the results shown in Fig.~\ref{fig_cusp}.
We have used smooth functions such as the standard four-fold 
anisotropy ($a_c(\theta) = 1 - \epsilon_4 \cos(4[\theta-\theta_R])$),
as well as a function with a deep but smooth Gaussian minimum,
$a_c(\theta) = 1 - \epsilon_c \exp((\theta-\theta_R)^2/w_c^2)$ 
with $\epsilon$ being the magnitude of the anisotropy and $w_c$ 
the width of the minimum. Note that, for these anisotropy functions, 
depending on the value of $\epsilon$, missing orientations
(signifying multiple solutions for a particular rotation angle)
can occur.

For smooth anisotropies, our results follow quite nicely the
prediction of the symmetric pattern approximation. When 
multi-valued solutions exist for a particular rotation 
angle -- one solution for a locked branch, one for an unlocked 
(zero angle) branch and one for an unstable branch (dotted in 
Fig.~\ref{fig_cusp}) -- the locked branch is well reproduced 
up to a certain angle ($\approx 30\deg$). For rotation angles
higher than this value, simulations always end up on the
unlocked branch. When the rotation angle is slowly increased
or decreased in small steps, the ``jump'' between the two branches
always occurs at the same angle. Therefore, multiple solutions
for a given rotation angle are never observed. This is in contrast
to results obtained with the boundary integral (BI) method \cite{Ghosh15}.
The reason for this discrepancy between the models is as of yet unknown.

\section{Grain competition}
For investigations of anisotropy in the azimuthal plane (the plane
perpendicular to the temperature gradient), three-dimensional
simulations are necessary. In bulk lamellar eutectics there are 
complex interactions between multiple eutectic grains and interfaces. 
In the absence of anisotropy, the microstructure consists of random 
lamellar patterns, because no particular orientation is favoured. 
When anisotropy is imparted in the system, the system begins to order 
depending on the underlying crystal structure of the solid. For 
example, if a two-fold anisotropy is present in the system, we 
obtain a perfectly lamellar array after a short transient. The
lamellae are always aligned with a minimum in the interface energy
function. When the latter is rotated with respect to the lateral
``walls'' (described by no-flux boundary conditions), the lamellae
are perpendicular to the boundaries close to the walls, but turn
to align with the minimum-energy direction over a distance of less
than one lamellar spacing.

\begin{figure}[h]
\centering
\begin{subfigure}[b]{0.35\textwidth}
\centering
\includegraphics[width=\textwidth]{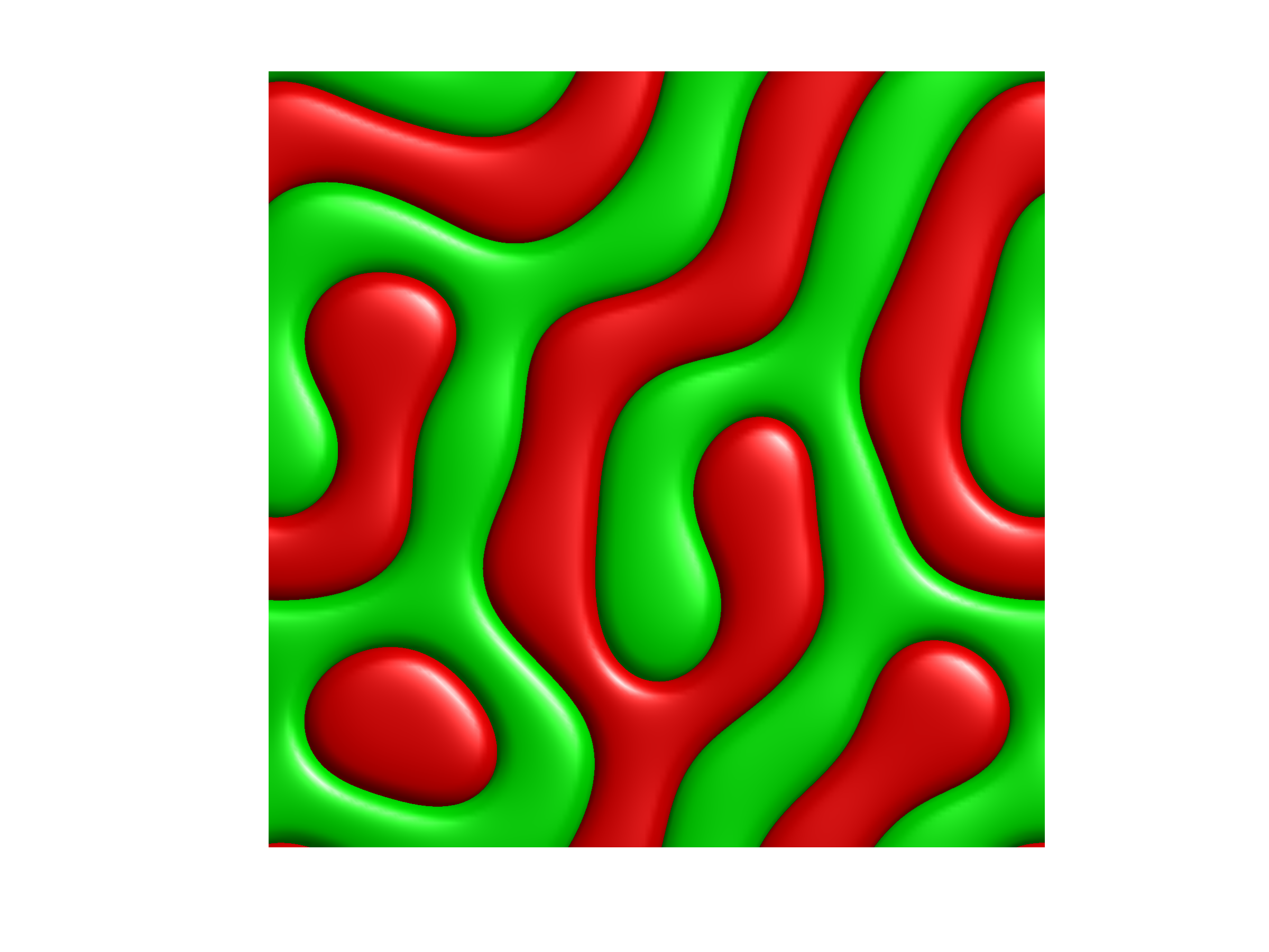}
\caption{$\tilde{t}$=54}\label{fig_random4}
\end{subfigure}%
\hspace*{-1cm}
\begin{subfigure}[b]{0.35\textwidth}
\centering
\includegraphics[width=\textwidth]{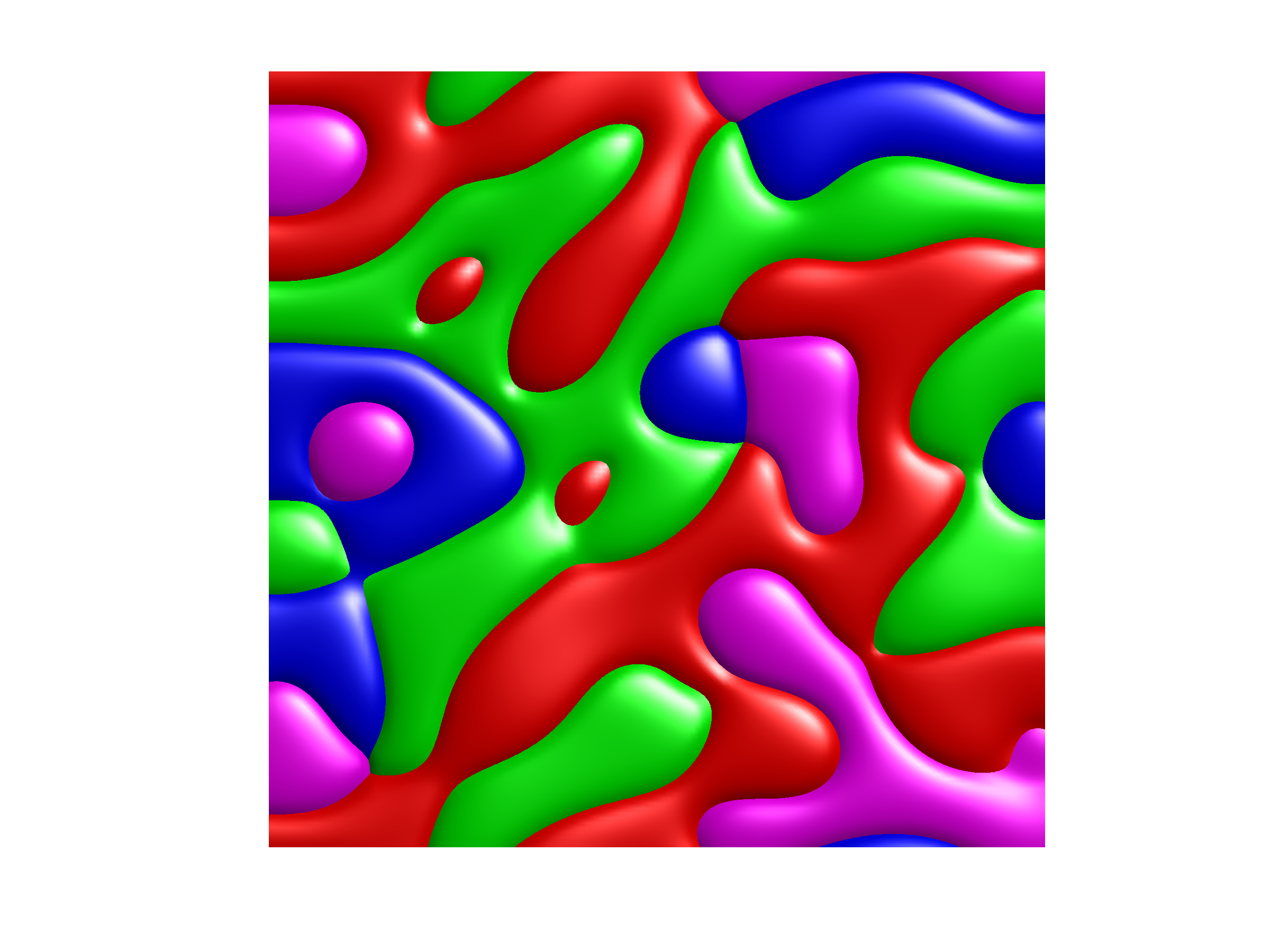}
\caption{$\tilde{t}$=9}\label{fig_img5}
\end{subfigure}%
\hspace*{-1cm}
\begin{subfigure}[b]{0.35\textwidth}
\centering
\includegraphics[width=\textwidth]{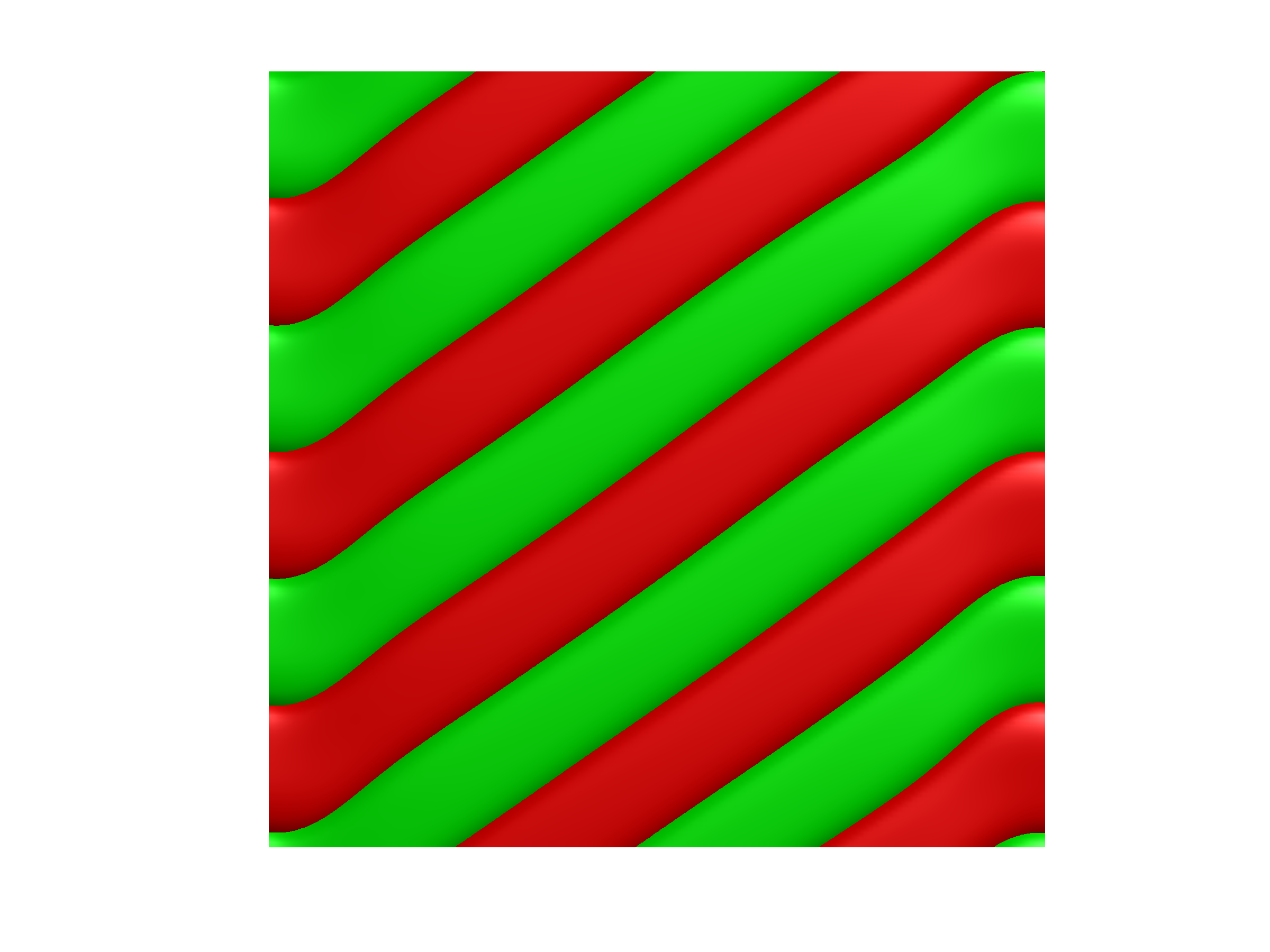}
\caption{$\tilde{t}$=54}\label{fig_alti}
\end{subfigure}
\caption{Top view of three-dimensional simulations. (a) Random lamellar pattern formation in a single eutectic grain in the absence of anisotropy. (b and c) time evolution of two eutectic grains: colors of $\alpha_1$, $\beta_1$, $\alpha_2$ and $\beta_2$ are red, green, blue and magenta, respectively. The IB between red and green phases are anisotropic, all others isotropic. The dimensionless time is $\tilde t=tV^2/D$}\label{fig_case1}
\end{figure}

In order to investigate the competition between eutectic grains,
we have considered the growth of two different eutectic grains
with different anisotropy functions $a_c$. In the example shown
in Fig.~\ref{fig_case1}, the interface energy between the ``red''
and ``green'' solids is 
given by $a_c = 1.0 + 0.3 \cos [2(\phi - \phi_R)]$, whereas
all other IB's are isotropic ($a_c=1$). The simulation is started
from a random tiling of the system with the various phases. 
A random lamellar pattern develops in the beginning, with a 
presence of local order in the system while there is no global 
order (Fig.~\ref{fig_img5}). Later on, lamellae of the red-green composite
start to grow along the preferred orientation. Finally, the system 
is left with only anisotropic interfaces, which are aligned with 
a minimum energy direction (Fig.~\ref{fig_alti}). In essence, surface 
tension anisotropy along the interphase boundaries induces an 
orientation relationship between the anisotropic phases yielding 
a regular lamellar array from an irregular eutectic structure.             

\section{Conclusions}
We have used phase-field simulations to investigate the influence
of crystallographic effects on eutectic solidification patterns.
We have found that an anisotropy of the solid-solid interphase 
boundaries can lead to important departures from the behavior
known for eutectics with isotropic interfaces. For an anisotropy
in a plane parallel to the temperature gradient, the competition
between the external temperature gradient and the IB energy
selects a growth direction that depends on the anisotropy strength
and differs from the direction of the temperature gradient, leading
to the growth of tilted lamellae. Anisotropy in the plane perpendicular
to the temperature gradient (azimuthal plane) leads to a selection
of particular orientations of the lamellae, which correspond to
minimum energy directions. This obviously favors the emergence
of ordered (parallel) lamellar arrays. A highly interesting
perspective is to understand this selection more quantitatively,
and to relate phase-field simulations to experiments on Al-Cu
alloys in which the crystallographic effects were characterized
in detail \cite{Hecht05,ulrike2014}.

\section*{Acknowledgments}
We thank Silv\`ere Akamatsu, Sabine Bottin-Rousseau, and Gabriel
Faivre for many stimulating discussions. This work was financially 
supported by Centre National d'\'Etudes Spatiales, France. 


\begin{thebibliography}{10}

\bibitem{Jackson66}
K.~A. Jackson and J.~D. Hunt.
\newblock Lamellar and rod eutectic growth.
\newblock {\em Transactions of the Metallurgical Society of AIME}, 236:1129,
  1966.

\bibitem{Hogan71}
LM~Hogan, RW~Kraft, and FD~Lemkey.
\newblock Eutectic grains.
\newblock {\em Adv. Mater. Res.}, 5:83--216, 1971.

\bibitem{Caroli92}
B~Caroli, C~Caroli, G~Faivre, and J~Mergy.
\newblock Lamellar eutectic growth of {$CBr_4-C_2Cl_6$}: effect of crystal
  anisotropy on lamellar orientations and wavelength dispersions.
\newblock {\em J. Cryst. Growth}, 118:135--150, 1992.

\bibitem{abhik2012}
A.~Choudhury and B.~Nestler.
\newblock Grand-potential formulation for multicomponent phase transformations
  combined with thin-interface asymptotics of the double-obstacle potential.
\newblock {\em Physical Review E}, 85:021602, 2012.

\bibitem{Ghosh15}
S.~Ghosh, A.~Choudhury, M.~Plapp, S.~Bottin-Rousseau, G.~Faivre, and
  S.~Akamatsu.
\newblock Intrephase anisotropy effects on lamellar eutectics: A numerical
  study.
\newblock {\em Physical Review E}, 91:022407, 2015.

\bibitem{silvere2012e}
S.~Akamatsu, S.~Bottin-Rousseau, M.~\c{S}erefo\u{g}lu, and G.~Faivre.
\newblock Lamellar eutectic growth with anisotropic interphase boundaries:
  Experimental study using rotational directional solidification.
\newblock {\em Acta Materialia}, 60:3206--3214, 2012.

\bibitem{silvere2012t}
S.~Akamatsu, S.~Bottin-Rousseau, M.~\c{S}erefo\u{g}lu, and G.~Faivre.
\newblock A theory of thin lamellar eutectic growth with anisotropic interphase
  boundaries.
\newblock {\em Acta Materialia}, 60:3199--3205, 2012.

\bibitem{Hoffman72}
DW~Hoffman and JW~Cahn.
\newblock Vector thermodynamics for anisotropic surfaces .1. fundamentals and
  application to plane surfcae junctions.
\newblock {\em Surface Science}, {31}:{368--\&}, {1972}.

\bibitem{Hecht05}
U~Hecht, VT~Witusiewicz, A~Drevermann, and S~Rex.
\newblock Orientation relationship in univariant {Al-Cu-Ag} eutectics.
\newblock {\em Trans. Indian Inst. Met.}, 58:545--551, 2005.

\bibitem{ulrike2014}
U.~Hecht, V.~Witisiewicz, and S.~Rex.
\newblock Solidification of bulk lamellar eutetcics.
\newblock {\em Materials Science Forum}, 790-791:343--348, 2014.

\end{thebibliography}

\end{document}